\documentclass[aps,prb,twocolumn,superscriptaddress,showpacs]{revtex4-1}
\usepackage{graphicx,bm,mathtools}

\usepackage{graphicx}
\usepackage{dcolumn}
\usepackage{bm}
\usepackage{xfrac}
\usepackage{booktabs}
\usepackage{threeparttable}

\begin{document}


\title{Evidence for two distinct superconducting phases in EuBiS$_2$F under pressure}
\author{C. Y. Guo}
\affiliation{Center for Correlated Matter and Department of Physics, Zhejiang University, Hangzhou 310058, China}
\author{Y. Chen}
\affiliation{Center for Correlated Matter and Department of Physics, Zhejiang University, Hangzhou 310058, China}
\author{M. Smidman}
\affiliation{Center for Correlated Matter and Department of Physics, Zhejiang University, Hangzhou 310058, China}
\author{S. A. Chen}
\affiliation{National Synchrotron Radiation Research Center, Hsinchu 30076, Taiwan}
\author{W. B. Jiang}
\affiliation{Center for Correlated Matter and Department of Physics, Zhejiang University, Hangzhou 310058, China}
\author{H. F. Zhai}
\affiliation{Center for Correlated Matter and Department of Physics, Zhejiang University, Hangzhou 310058, China}
\author{Y. F. Wang}
\affiliation{Center for Correlated Matter and Department of Physics, Zhejiang University, Hangzhou 310058, China}
\author{G. H. Cao}
\affiliation{Center for Correlated Matter and Department of Physics, Zhejiang University, Hangzhou 310058, China}
\affiliation{Collaborative Innovation Center of Advanced Microstructures, Nanjing 210093, China}
\author{J. M. Chen}
\affiliation{National Synchrotron Radiation Research Center, Hsinchu 30076, Taiwan}
\author{X. Lu}
\email{xinluphy@zju.edu.cn}
\affiliation{Center for Correlated Matter and Department of Physics, Zhejiang University, Hangzhou 310058, China}
\affiliation{Collaborative Innovation Center of Advanced Microstructures, Nanjing 210093, China}
\author{H. Q. Yuan}
\email{hqyuan@zju.edu.cn}
\affiliation{Center for Correlated Matter and Department of Physics, Zhejiang University, Hangzhou 310058, China}
\affiliation{Collaborative Innovation Center of Advanced Microstructures, Nanjing 210093, China}

\begin{abstract}

 We present a pressure study of the electrical resistivity, ac magnetic susceptibility and powder x-ray diffraction (XRD) of the recently discovered BiS$_2$-based superconductor EuBiS$_2$F. At ambient pressure, EuBiS$_2$F shows an anomaly in the resistivity at around $T_0\approx~280$ K and a superconducting transition at $T_c\approx~0.3$ K. Upon applying hydrostatic pressure, there is little change in $T_0$ but the amplitude of the resistive anomaly is suppressed, whereas there is a dramatic enhancement of $T_c$ from 0.3~K to about 8.6~K at a critical pressure of $p_c$ $\approx{1.4}$~GPa. XRD measurements confirm that this enhancement of $T_c$ coincides with a structural phase transition from a tetragonal phase ($P4/nmm$) to a monoclinic phase ($P2_1$/m), which is similar to that observed in isostructural LaO$_{0.5}$F$_{0.5}$BiS$_2$. Our results suggest the presence of two different superconducting phases with distinct crystal structures in EuBiS$_2$F, which may be a general property of this family of BiS$_2$-based superconductors.
\end{abstract}


\maketitle
\section{Introduction}

The recently discovered BiS$_2$-based superconductors\cite{BOS, LaBS1} have attracted considerable attention. Similar to cuprate and iron-pnictide superconductors,\cite{cupr,Fe} these compounds possess layered crystal structures but with a rather low superconducting transition temperature ($T_c$). The first BiS$_2$-based superconductor Bi$_4$O$_4$S$_3$, which crystalizes in a body centered tetragonal structure (space group $I4/mmm$) consisting of double layers of BiS$_2$, separated by units of Bi$_2$O$_2$ and SO$_4$, shows a metallic normal state before becoming superconducting at $T_c$~$\approx$~5~K.\cite{BOS} Subsequently, another family of BiS$_2$-based superconductors $L$O$_{1-x}$F$_{x}$BiS$_2$($L$~=~La, Ce, Pr or Nd)\cite{LaBS1,CeBS,PrBS2,NdBS,NdBS2} with a primitive tetragonal structure (space group $P4/nmm$) has also been studied intensively. This structure contains the same double BiS$_2$ layers but in this instance they are separated by units of $L$O$_{1-x}$F$_x$. At ambient pressure, these compounds become superconducting upon adding charge carriers via substituting F for O and there is a change from insulating to semiconducting behavior.

A dramatic enhancement of $T_c$ was observed on applying pressure to $L$O$_{1-x}$F$_{x}$BiS$_2$($L$~=~La, Ce, Pr or Nd),\cite{Wolo PRB,Tomita La,Wolo JPCM} which seems to be a general feature of this family of BiS$_2$ based superconductors. For example, LaO$_{0.5}$F$_{0.5}$BiS$_2$ becomes superconducting at $T_c\simeq 2.5~$K at ambient pressure, but its $T_c$ undergoes a sharp increase at $p\simeq$0.7~GPa, reaching $T_c\simeq$~10.7~K.\cite{Tomita La} Furthermore, a similarly large $T_c$ was also measured at ambient pressure for LaO$_{0.5}$F$_{0.5}$BiS$_2$ samples synthesized in high pressure conditions.\cite{La HP syn} However, the underlying mechanism behind the pressure-induced enhancement of $T_c$ remains unresolved. The increase of $T_c$ under pressure could be due to modifications of the electronic structure and in particular, it was suggested that it may be related to an increase in the density of charge carriers upon the suppression of semiconducting behavior.\cite{Wolo JPCM}  Differences in the Hall effect measurements of Eu$_3$Bi$_2$S$_4$F$_4$  at low- and high- pressures  also suggest a pressure induced change in the electronic structure.\cite{Eu3P} It was also found that in LaO$_{0.5}$F$_{0.5}$BiS$_2$, the jump in $T_c$ is accompanied by a structural phase transition from a tetragonal to a monoclinic structure.\cite{Tomita La} On the other hand, different behavior is shown in Bi$_4$O$_4$S$_3$, where $T_c$ decreases upon applying pressure, reaching about 3~K at 2~GPa,\cite{BOS p} and the material is metallic in the normal state. The parent compound Bi$_6$O$_8$S$_5$ is a band insulator, but a 50\% deficiency of SO$_4$ generates additional electron carriers within the BiS$_2$ layers of Bi$_4$O$_4$S$_3$, leading to metallic behavior.\cite{review} Therefore, one may expect that the sudden increase of $T_c$ under pressure in $L$O$_{1-x}$F$_{x}$BiS$_2$ is associated with the suppression of a semiconducting gap and it would not be observed in metallic systems.

The newly discovered BiS$_2$-based superconductor EuBiS$_2$F, which is isostructural to $L$O$_{1-x}$F$_{x}$BiS$_2$, displays unique properties.\cite{EFBS} It not only becomes superconducting at $T_c$ $\approx$~0.3 K, but there is also an anomaly in the resistivity at around 280 K, which has been proposed to be a charge density wave (CDW) transition.\cite{EFBS} There is no change in the crystal structure down to 13~K but a small anomaly in the ratio of the lattice parameters $c/a$ is observed around 280~K. The strong hybridization between Eu-$4f$ and Bi-$6p$ electrons results in a large electronic specific heat coefficient of about 73~mJ/mol~K$^2$. Importantly, this compound does not exhibit significant semiconducting behavior at low temperatures, which is different from most other $L$O$_{1-x}$F$_{x}$BiS$_2$ superconductors. Therefore, EuBiS$_2$F may provide us a valuable opportunity to study the origin of the dramatic enhancement of $T_c$ in $L$O$_{1-x}$F$_{x}$BiS$_2$ under pressure and also the possible interplay of superconductivity and CDW order. In this paper, we report electrical resistivity, ac magnetic susceptibility and powder x-ray diffraction measurements of EuBiS$_2$F under high pressure. It is found that the resistive anomaly around 280~K is weakened under pressure and vanishes around $p_c$ $\approx{1.4}$~GPa. On the other hand, the superconducting transition temperature undergoes an enormous increase from $T_c\simeq~0.3$~K to 8.6~K at $p_c$, where a structural phase transition from the tetragonal phase ($P4/nmm$) to a monoclinic phase ($P2_1$/m) takes place. These results suggest that two distinct superconducting phases, corresponding to two distinct crystal structures, exist in EuBiS$_2$F.

\section{Experimental Methods}

Polycrystalline samples of EuBiS$_2$F were synthesized by a solid-state reaction method, as described in Ref.~\onlinecite{EFBS}. Resistivity measurements under hydrostatic pressure were performed up to about 2.4~GPa using a piston-cylinder-type pressure cell, with Daphne 7373 used as a pressure transmitting medium. The applied pressure was determined from measuring the shift in $T_c$ of a high quality Pb sample. High temperature measurements were performed using a Physical Property Measurement System (Quantum Design PPMS-14T) in a temperature range of 2 to 300 K. The low temperature measurements were performed using a $^3$He refrigerator down to 0.3~K. All resistivity measurements were carried out using a standard four-probe method with four Pt wires spot-welded to the sample surface. ac magnetic susceptibility measurements were carried out with a set of coils designed in-house, consisting of a drive coil, a pick-up coil and a coil for compensation. The system was driven with an applied current of 0.1~mA at a frequency of 1523~Hz and the voltage signal was measured using a SR-830 lock-in amplifier. The dc magnetic susceptibility was measured using a superconducting quantum interference device (SQUID) magnetometer, Magnetic Property Measurement System (Quantum Design MPMS-5T). High pressure powder x-ray diffraction (XRD) measurements were carried out at the BL12B2 Taiwan beamline of Spring-8. Finely grained EuBiS$_2$F powder and spread-out tiny ruby balls for pressure determination were filled into the pinhole (diameter 235~$\mu$m) of a stainless steel gasket mounted on a diamond anvil cell with a culet size of 450~$\mu$m. A mixture of methanol, ethanol and water in a ratio of 16:3:1 was used as the pressure-transmitting medium in the XRD measurements.\cite{PMed} Measurements of the fluorescence line shift were performed at multiple positions in the sample chamber using a laser with a diameter of 5~-~10~$\mu$m. This was measured before and after each exposure to confirm the hydrostatic nature of the applied pressure. With a monochromatic beam ($\lambda$ = 0.68969~\AA), the 2-D diffraction images were collected using an ADSC Quantum 4R CCD x-ray detector and then transformed into 1-D patterns by using the program FIT2D. A high quality CeO$_2$ standard (99.99\%, Aldrich) was used to calibrate the experimental setup and determine the sample-to-detector distance.

\section{Results}
\subsection{Resistivity and ac susceptibility}

\begin{figure}[t]\centering
  \includegraphics[width=0.8\columnwidth]{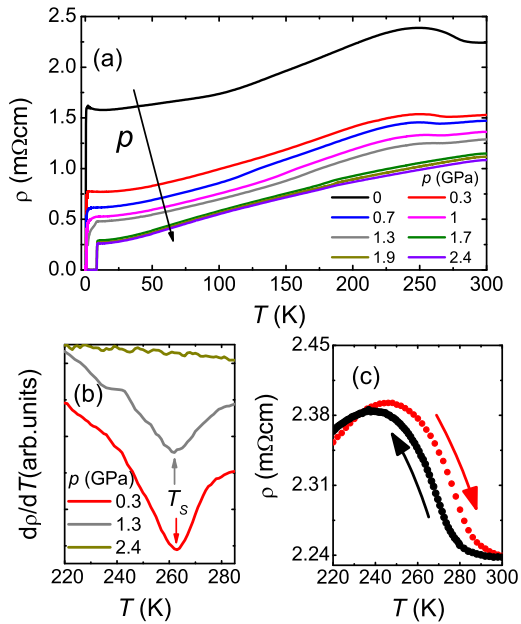}
\caption{(Color online) (a) Temperature dependence of the electrical resistivity $\rho(T)$ of EuBiS$_2$F at various pressures in the temperature range 0.3  to 300~K. (b) The derivative of $\rho(T)$  across the high temperature region, where the presence of the anomaly at $T_0$ is shown by the minimum of the data. (c) The resistivity anomaly at $T_0$ ($p=0$), measured upon both heating and cooling which shows clear hysteresis.}\label{Fig1}
\end{figure}

\begin{figure}[tb]\centering
  \includegraphics[width=0.8\columnwidth,clip=true, trim= 67 8 135 45]{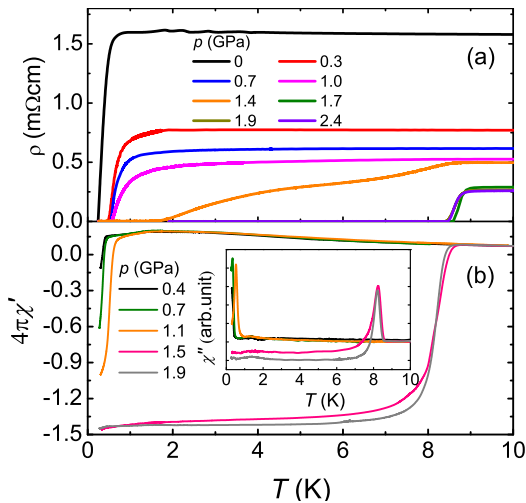}
\caption{(Color online) (a) Temperature dependence of the electrical resistivity $\rho(T)$ of EuBiS$_2$F below 10~K, in the vicinity of the superconducting transition. (b) Temperature dependence of the real part of the ac magnetic susceptibility $\chi'$ measured at several pressures, with the imaginary component $\chi''$ shown in the inset}\label{Fig2}
\end{figure}

The temperature dependence of the electrical resistivity [$\rho(T)$] of EuBiS$_2$F at various pressures is displayed in Fig.~\ref{Fig1}(a). At ambient pressure, $\rho(T)$ shows a broad hump below $\simeq 280$ K, which has been proposed to be a CDW-type transition.\cite{EFBS} At low temperatures, a superconducting transition is observed with $T_c\simeq$~0.3~K, which is consistent with previous measurements. Figure~\ref{Fig1}(b) displays the derivative of $\rho(T)$ in the high temperature region where the minimum at $T_0$ corresponds to the broad hump. The temperature where this feature occurs changes little with increasing pressure up to 1.3~GPa, but the amplitude weakens. For pressures greater than 1.3~GPa, the anomaly can not be detected and the resistivity monotonically decreases below 300~K. The high temperature resistivity is shown in Fig.~\ref{Fig1}(c), measured upon heating and cooling. The difference between the curves clearly shows the presence of hysteresis associated with the anomaly at $T_0$.

Unlike the possible CDW transition at $T_0$, the superconducting transition of EuBiS$_2$F varies considerably upon applying pressure. As shown in Fig.~\ref{Fig2}(a), for $p<p_c\simeq1.4$ GPa there is a slight increase of $T_c$ with increasing pressure and the superconducting transition is broader. Upon increasing the pressure to 1.7~GPa, $T_c$ significantly increases from below 1~K to around 8.6~K. Further increasing the pressure above 1.7~GPa results in no significant change of $T_c$ and the superconducting transition remains sharp. At the critical pressure of $p_c\simeq 1.4$~GPa, the superconducting transition becomes significantly broader, which is likely attributable to small pressure inhomogeneities. In addition, the pressure dependence of the residual resistivity in the normal state also exhibits a distinct change between 1.4 and 1.7~GPa. Below 1.4~GPa, the residual resistivity decreases upon applying pressure, but at higher pressures, there is no significant change. Measurements of the real part of the ac magnetic susceptibility ($\chi'$) under pressure are shown in  Fig.~\ref{Fig2}(b). The data are given in absolute units which were obtained by scaling $\chi'(T)$ above 2~K to the dc magnetic susceptibility measured in the MPMS. At a low pressure of $p~=~0.4$~GPa, the onset of diamagnetic shielding occurs at 0.4~K, which is consistent with the $T_c$ observed in resistivity measurements. There is a slight increase in $T_c$ with pressure, reaching 0.7~K at 1.1~GPa. Due to the low $T_c$ and broadening of the transition, $\chi'$ continues to decrease even at the lowest measured temperature of 0.3~K. At a pressure of 1.5~GPa, there is a  large enhancement of $T_c$ to 8.6~K, which is in good agreement with the resistivity measurements. At low temperatures $\chi'$ reaches a near constant value of about -1.4, which is lower than the value of -1 corresponding to full diamagnetic shielding and this is likely to be due to demagnetization effects. The imaginary part of the ac magnetic susceptibility $\chi''$ is shown in the inset and the peak is due to energy loss at the onset of superconductivity.

\begin{figure}[t]\centering
\includegraphics[width=0.85\columnwidth]{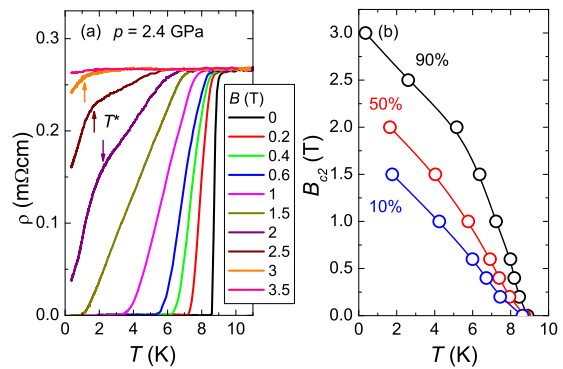}
\caption{(Color online) Temperature dependence of the resistivity of EuBiS$_2$F under applied magnetic fields at about 2.4~GPa. Panel (a) displays temperature dependence of the resistivity at various magnetic fields, while panel (b) shows the upper critical field determined from the temperature where the resistivity drops to 90\%, 50\% and 10\% of the normal state value ($\rho_n$).}
\label{Fig3}
\end{figure}

Resistivity measurements were also performed at the highest pressure ($p~\approx$~2.4~GPa) with an applied magnetic field. It can be seen in Fig.~\ref{Fig3}(a) that with increasing magnetic field, the superconducting transition is shifted to lower temperatures and becomes significantly broader. The results at high fields show a kink at $T^{*}$, which is similar to that observed in the high pressure phase of polycrystalline Eu$_3$Bi$_2$S$_4$F$_4$.\cite{Eu3P} As EuBiS$_2$F has a layered structure, its upper critical field is expected to be anisotropic, as observed in isostructural LaO$_{0.5}$F$_{0.5}$BiS$_2$.\cite{La M} Since a polycrystalline sample contains grains of different orientations, the upper critical field may vary from grain to grain, leading to a kink in the resistivity at high magnetic fields. However, we also cannot exclude other possibilities at the present time, such as  a field-induced transition inside the superconducting state. Figure~\ref{Fig3}(b) shows the upper critical field ($B_{c2}$) determined from where the resistivity reaches 90\%, 50\% and 10\% of the normal state value just above $T_c$ ($\rho_n$). The value at zero temperature $B_{c2}$(0) was determined to be approximately 3.1~T or 1.8~T from extrapolating the 90\%$\rho_n$ or 10\%$\rho_n$ curves respectively to $T=0$. This corresponds to respective Ginzburg-Landau coherence lengths of $\xi_{\rm GL}\simeq 103$~\AA $~$and $\simeq 135$~\AA. The value of $B_{c2}(0)/T_c$ for the 90\%$\rho_n$ data decreases from 0.4 T/K ($p=0$) \cite{EFBS} to 0.35 T/K ($p~\approx~2.4$~GPa). A similar value is found in Sr$_{0.5}$La$_{0.5}$FBiS$_2$ at ambient pressure,\cite{SrBiS2} while a larger value is estimated in LaO$_{0.5}$F$_{0.5}$BiS$_2$ synthesized at high pressure.\cite{LaBS1}

\subsection{X-ray diffraction}

\begin{figure}[t]\centering
\includegraphics[width=0.8\columnwidth]{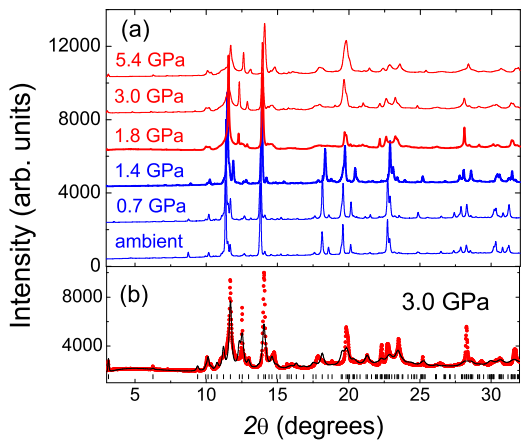}
\caption{(Color online) (a) Room temperature powder x-ray diffraction patterns of polycrystalline EuBiS$_2$F at various pressures. Additional Bragg peaks are observed for $p>p_c\simeq 1.4$~GPa, evidencing a structural phase transition. (b) XRD pattern at $p~=~3.0$~GPa and its refinement for a monoclinic crystal structure with space group $P2_1/m$. }
\label{Fig4}
\end{figure}

To examine whether there is a change in crystal structure correlated with the enhancement of $T_c$ in EuBiS$_2$F, we performed  high pressure powder XRD measurements at room temperature, as shown in Fig.~\ref{Fig4}(a). At ambient pressure, EuBiS$_2$F crystallizes in a tetragonal structure with the space group $P4/nmm$, which is consistent with the previous report.\cite{EFBS} It can be seen that all the patterns measured at applied pressures display an additional peak at around 28.1$^\circ$ due to diffraction from the ruby balls, which are used to determine the pressure in the diamond anvil cell. At applied pressures greater than 1.4~GPa, different Bragg peaks are observed, which is clear evidence for a change in crystal structure and an increase in the number of peaks indicates a reduction in the crystal symmetry. Similar behavior is observed in LaO$_{0.5}$F$_{0.5}$BiS$_2$,\cite{Tomita La} where a structural phase transition occurs from a tetragonal ($P4/nmm$) to monoclinic($P2_1/m$) structure. Upon further increasing pressure, no additional changes in the crystal structure are observed up to 12.6~GPa. To determine the crystal structure at high pressures, we analyzed the diffraction pattern at 3.0~GPa which is displayed in Fig.~\ref{Fig4}(b). The pattern was fitted by performing a structural refinement using the GSAS+EXPGUI software,\cite{GSAS} with the monoclinic structure from Ref.~\onlinecite{Tomita La}. The solid line shows the fitted curve and this indicates that the Bragg peaks at high pressure can be reasonably accounted for by this structure. It should be noted that although there is a good agreement between the theoretical and experimental peak positions, the peak intensities are not perfectly matched. Inaccuracies of the relative peak intensities are a frequent problem when measuring powder XRD in a diamond anvil cell due to the small sample mass and the presence of preferred orientations.\cite{prefO} Therefore, the atomic positions of the pressure induced phase may be best determined from density functional theory calculations as reported previously,\cite{Tomita La} rather than from refinements of XRD data.

\section{Summary and discussion}

\begin{figure}[t]\centering
\includegraphics[width=0.8\columnwidth]{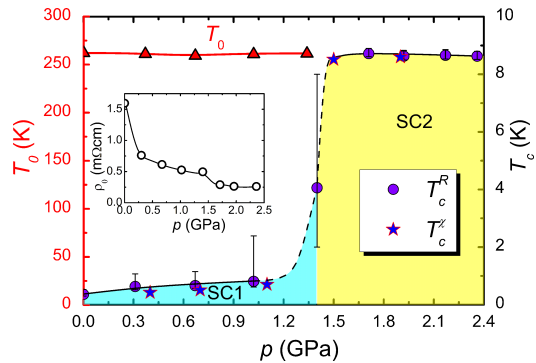}
\caption{(Color online)Temperature-pressure phase diagram for EuBiS$_2$F in which both $T_c$ and $T_0$ are displayed. Values of $T_c^R$ were taken from the mid-point of the resistivity transitions, while $T_c^{\chi}$ are from the onset of the transition in the ac magnetic susceptibility; the two experimental methods give consistent results. The uncertainties in $T_c^R$ were estimated from where the resistivity drops to 10\% and 90\% of the normal state value. The inset shows the change of the residual resistivity in the normal state ($\rho_0$) with applied pressure. }
\label{Fig5}
\end{figure}

The results measured under applied pressure are summarized by the phase diagram in Fig.~\ref{Fig5}. Below $p_c\simeq$~1.4~GPa, $T_c$ increases slightly with increasing pressure. On the other hand, the position of the broad hump in the resistivity at $T_0$ is nearly unchanged, but its amplitude is reduced with increasing pressure. Around $p_c\simeq1.4$ GPa, there is a dramatic enhancement of $T_c$, whereas the high temperature resistive anomaly at $T_0$ vanishes. The large uncertainty of $T_c$ at 1.4~GPa reflects the broad nature of the transition, arising from the coexistence of the low- and high-pressure phases as a result of the small pressure inhomogeneities in the measurements. In addition, the crystal structure undergoes a transition from a tetragonal ($P4/nmm$) to a monoclinic($P2_1/m$) structure between 1.4~GPa and 1.8~GPa and the residual resistivity also shows a step-like decrease across the critical pressure region of $p_c\simeq~1.4$~GPa. Upon further increasing pressure, there is no significant change of $T_c$.

The above phase diagram of EuBiS$_2$F is similar to that observed in other $L$O$_{1-x}$F$_{x}$BiS$_2$ superconductors, with all the compounds showing a comparable value of $T_c$ at high pressures. In contrast to other materials, EuBiS$_2$F does not show significant semiconducting behavior and therefore the enhancement of $T_c$ is not simply associated with the suppression of this behavior. However, the high-pressure XRD results reveal a similar structural phase transition in both LaO$_{0.5}$F$_{0.5}$BiS$_2$ \cite{Tomita La} and EuBiS$_2$F (this study). This suggests that two different superconducting phases, associated with two distinct crystal structures, may exist in $L$O$_{1-x}$F$_{x}$BiS$_2$; the low-$T_c$ superconducting phase (SC1) in the tetragonal structure and the high-$T_c$ phase (SC2) in the monoclinic structure as shown in Fig.~\ref{Fig5}. The dramatic enhancement of $T_c$ under pressure is likely due to differences in the electronic structure between these two structural phases which could be examined by band-structure calculations. Furthermore, it is also important to extend the high-pressure XRD measurements to other compounds in order to confirm the generality of this observation. In comparison with other $L$O$_{1-x}$F$_{x}$BiS$_2$ superconductors, EuBiS$_2$F shows a greatly reduced $T_c$ at low pressures which could be attributable to the Eu-$4f$ magnetic moments or its valence state.\cite{EFBS} However, the values of $T_c$ for all these compounds are similar in the high-pressure region \cite{Wolo JPCM,Wolo PRB} and therefore the structural phase transition at $p_c\simeq~1.4$~GPa is likely accompanied by a change of the Eu-valence, which may also lead to the disappearance of the resistive anomaly at $T_0$ above $p_c$. To confirm the possible changes of the Eu-valence in EuBiS$_2$F under pressure, it would be desirable to perform experiments such as x-ray absorption spectroscopy.

In summary, we have performed electrical resistivity, ac magnetic susceptibility and XRD measurements of EuBiS$_2$F under pressure.  It is found that $T_c$ of EuBiS$_2$F is dramatically enhanced at $p_c\simeq~1.4$~GPa, around which pressure the resistive anomaly at $T_0\simeq~280$~K vanishes and a structural phase transition is observed at room temperature. Our results suggest that EuBiS$_2$F has two different superconducting phases corresponding to two different crystal structures; a low-$T_c$ superconducting phase in the tetragonal structure and a high-$T_c$ phase in the monoclinic structure. These observations may be a general feature of this family of BiS$_2$-based superconductors, which needs to be confirmed by further experiments. Furthermore, this also implies that the monoclinic structure may favor a superconducting phase with a higher $T_c$, providing additional insights towards how $T_c$ may be enhanced.

\section*{Acknowledgments}

We thank C. Cao for useful discussions. This work was supported by the National Basic Research Program of China (No. 2011CBA00103), the National Natural Science Foundation of China (No.11474251 and No.11174245) and the Fundamental Research Funds for the Central Universities.


\begin{thebibliography}{10}

\bibitem{BOS} Y. Mizuguchi, H. Fujihisa, Y. Gotoh, K. Suzuki, H. Usui, K. Kuroki, S. Demura, Y. Takano, H. Izawa, and O. Miura, Phys. Rev. B \textbf{86}, 220510(R) (2012).
\bibitem{LaBS1} Y. Mizuguchi, S. Demura, K. Deguchi, Y. Takano, H. Fujihisa, Y. Gotoh, H. Izawa,
O. Miura,  J. Phys. Soc. Jpn. \textbf{81} 114725 (2012).
\bibitem{cupr} J. G. Bednorz and K. A. Muller, Z. Phys. B \textbf{64}, 189 (1986).
\bibitem{Fe} Y. Kamihara, T. Watanabe, M. Hirano, and H. Hosono, J. Am. Chem. Soc. \textbf{130}, 3296 (2008).
\bibitem{CeBS} J. Xing, S. Li, X. Ding, H. Yang, and H. H. Wen, Phys. Rev. B \textbf{86}, 214518 (2012).
\bibitem{PrBS2} R. Jha, A. Kumar, S. K. Singh, and V. P. S. Awana, J. Supercond. Novel Magn. \textbf{26}, 499 (2013).
\bibitem{NdBS}S. Demura, Y. Mizuguchi, K. Deguchi, H. Okazaki, H. Hara, T. Watanabe, S. J. Denholme, M. Fujioka, T. Ozaki, H. Fujihisa, Y. Gotoh, O. Miura, T. Yamaguchi, H. Takeya, and Y. Takano, J. Phys. Soc. Jpn. \textbf{82}, 033708 (2013).
\bibitem{NdBS2} R. Jha, A. Kumar, S. K. Singh, and V. P. S. Awana, J. Appl. Phys. \textbf{113}, 056102 (2013).
\bibitem{Wolo PRB} C. T. Wolowiec, D. Yazici, B. D. White, K. Huang, and M. B. Maple, Phys. Rev. B \textbf{88}, 064503 (2013).
\bibitem{Tomita La} T. Tomita, M. Ebata, H. Soeda, H. Takahashi, H. Fujihisa, Y. Gotoh, Y. Mizuguchi, H. Izawa, O. Miura, S. Demura, K. Deguchi, and Y. Takano, J. Phys. Soc. Jpn. \textbf{83}, 063704 (2014).
\bibitem{Wolo JPCM} C. T. Wolowiec, B. D. White, I. Jeon, D. Yazici, K. Huang, and M. B. Maple, J. Phys.: Condens. Matter \textbf{25}, 422201 (2013).
\bibitem{La HP syn} K. Deguchi, Y. Mizuguchi, S. Demura, H. Hara, T. Watanabe, S. J. Denholme, M. Fujioka, H. Okazaki, T. Ozaki, H. Takeya, T. Yamaguchi, O. Miura, and Y. Takano, Europhys. Lett. \textbf{101}, 17004 (2013).
\bibitem{Eu3P} Y. K. Luo, H. F. Zhai, P. Zhang, Z. A. Xu, G. H. Cao, and J. D. Thompson, Phys. Rev. B \textbf{90}, 220510(R) (2014).
\bibitem{BOS p} H. Kotegawa, Y. Tomita, H. Tou, H. Izawa, Y. Mizuguchi, O. Miura, S. Demura, K. Deguchi, and Y. Takano,J. Phys. Soc. Jpn. \textbf{81}, 103702 (2012).
\bibitem{review} Y. Mizuguchi, J. Phys. Chem. Solids \textbf{84,} 34 (2015).
\bibitem{EFBS} H. F. Zhai, Z. T. Tang, H. Jiang, K. Xu, K. Zhang, P. Zhang, J. K. Bao, Y. L. Sun, W. H. Jiao, I. Nowik, I. Felner, Y. K. Li, X. F. Xu, Q. Tao, C. M. Feng, Z. A. Xu, and G. H. Cao, Phys. Rev. B \textbf{90}, 064518 (2014).
\bibitem{PMed} R. J. Angel, M. Bujak, J. Zhao, G. D. Gatta and S. D. Jacobsen, J. Appl. Crystallogr. \textbf{40}, 26-32 (2007).
\bibitem{La M} M. Nagao, A. Miura, S. Demura, K. Deguchi, S. Watauchi, T. Takei, Y. Takano, N. Kumada, and I. Tanaka, Solid State Commun. \textbf{178}, 33 (2014).
\bibitem{SrBiS2} X. Lin, X. X. Ni, B. Chen, X. F. Xu, X. X. Yang, J. H. Dai, Y. K. Li,  X. J. Yang, Y. K. Luo, Q. Tao, G. H. Cao, and Z. Xu, Phys. Rev. B \textbf{87}, 020504(R) (2013).
\bibitem{GSAS} B. H. Toby, J. Appl. Crystallogr. \textbf{34}, 210-213  (2001).
\bibitem{prefO} O. Tschauner, J. McClure, and M. Nicol, J. Synchrotron Rad. \textbf{12}, 626-631 (2005).


\end{thebibliography}
\end{document}